\documentstyle[12pt]{article}
\begin{document}

\centerline{\large\bf STUDY OF A PLANAR LATTICE MODEL}
\vskip .1in
\centerline{\large\bf WITH P$_4$ INTERACTION}
\vskip 0.5in

\centerline{Kisor Mukhopadhyay, Abhijit Pal and Soumen Kumar Roy$^1$}

\vskip 0.2in

\centerline{Department of Physics, Jadavpur University,}
\centerline{Calcutta 700032, INDIA}
\vskip 0.5in

\begin{abstract}
  A planar square lattice model with 3-d spins interacting with nearest 
neighbours through a potential -$\epsilon P_4 (cos \theta_{ij})$ is
studied by Monte Carlo technique. Lattice sizes from 10$\times$10 to  
30$\times$30 are considered for calculating various thermodynamic averages. 
A 80$\times$80 lattice has been used to obtain the pair correlation function. 
To accurately ascertain the order of the phase transition the 
Ferrenberg-Swendsen technique has been used on a 120$\times$120 lattice. Our 
study predicts that the system exhibits a first order phase transition which 
is confirmed by the twin-peaked nature of the distribution function. The pair
correlation function shows an algebraic decay at low temperatures and an 
exponential decay at high temperatures. Mean field and Two-site Cluster 
calculations have also been performed and the latter is found to predict 
the thermodynamic averages fairly accurately.  
\end{abstract}
\vskip .3in
{\bf PACS number(s) :} 61.30., 61.30. Gd., 64.70. M. 

\footnotetext[1]{Corresponding author :\\
\indent E-mail : skroy@juphys.ernet.in}
\clearpage
\indent
The phenomenon of fluctuation destruction of long range order in two 
dimensional systems with a continuous symmetry is well known in literature 
and is referred to as the Mermin-Wagner-Berezinski theorem. Thus there
 can be no true long range order in such systems. Systems with power law decay 
of order parameter correlation function are said to have $\:$quasi-long-range 
$\:$order
(QLRO). The $\:$much studied $\:$XY-model $\:$has a $\:$Hamiltonian $\:$$H 
= -\epsilon \sum_{x,\mu} 
(\sigma (x), \sigma (x+\mu))$ where $\sigma(x)$ is a two component unit vector 
at the
site $x$ of a square lattice, $\mu$ denotes the two directions of the lattice 
and 
$(\sigma (x), \sigma (y))$ denotes the scalar product of the spin vectors 
$\sigma (x)$
and $\sigma (y)$. There are topological point defects in the two dimensional 
XY model 
and vortex unbinding leads to a QLRO-disorder transition which is second order
 [1].

The question that naturally arises is whether systems with d=2, n$\ge$2 
(where 
d is the system dimensionality and n is the spin dimensionality) has 
topological 
phase transitions. In the class of O(3) Heisenberg system Lau and Dasgupta 
has shown 
in a numerical study [2] that the phase transition disappears in absence of 
point defects 
and the system remains in its low temperature ordered phase.

Another class of O(3) system of interest is where the Hamiltonian is 
\begin{equation}
H = -\epsilon \sum_{i,j} P_L (cos \gamma_{i,j})
\end{equation} 
where $i$,$j$ are nearest neighbours and $cos (\gamma_{i,j}) = 
(\sigma_{i},\sigma_{j})$. Chiccoli, Pasini and Zannoni [3] have numerically 
investigated this system
for L=2 and found that the order parameter correlation function exhibits a 
power
law decay in the ordered phase and an exponential decay in the high 
temperature phase. A far more elaborate 
study in this system (and also in d=2, n=40 model) has more recently
 been made by Kunz and
Zumbach [4]. These authors have found strong evidence for a topological phase 
transition driven
by defects. In the n=3 case, this transition is associated with a divergence of
correlation length and a cusp in the specific heat.

In this Letter we present a Monte Carlo study on a square lattice with 
3-dimensional spins 
interacting with nearest neighbours through the Hamiltonian (1) for L=4. 
The 
energy and order parameters $<P_2>$ and $<P_4>$ have been computed for 
10$\times$10, 
20$\times$20 and 30$\times$30 systems. We have chosen a 80$\times$80 
lattice to
study the decay of the order parameter correlation function and   
 have found an algebric decay 
 in the low temperature phase and an exponential decay at high temperatures. 
 To accurately ascertain the 
order of the phase transition we have employed the Ferrenberg-Swendsen 
reweighting
technique [5] on a 120$\times$120 lattice. 
The study on the 120$\times$120 system shows that transition is of first 
order nature 
with the probability distribution function exhibiting a double peak 
structure.

We start with calculations based on the mean field (MF) theory and the two
 site cluster (TSC) theory
on these systems. This may seem to be of  academic interest only because in a 
 strongly fluctuating two dimensional system the MF theory is not expected to 
perform well, but the TSC calculations, as will be seen, succed fairly well in 
explaining the MC results.  

The MF approximation [6,7] can be developed for the models defined by the
equation (2). In the MF approximation scheme, the singlet orientational
distribution function is given by \\
\begin{equation}
 P\,({\rm cos}\theta)\,=\,\frac{exp\left[z\, \beta\, S\, P_{L}\,({\rm
cos}\theta)\right]}{Z_1}  
\end{equation}
where
\begin{equation}
  Z_1 \,=\,\int_{0}^{\pi}d\theta \, sin \theta \,
exp\left[(z/T^\ast) \, S \, P_L(cos \theta) \right] 
\end{equation}
is the single-particle pseudo-partition function with $L$=4 for 
the present model and $\theta$ is the
orientation of the particle axis with respect to the director. Here z is
the lattice coordination number (z=4 in the present case) and S is some
variational parameter. The difference in the free energy per particle
between the ordered and the isotropic phases is given by \\
\begin{equation}
 \beta\, F_{0}^{\ast}\,=\,(z/2)\, \beta\, S^2\,-\,log\, Z_1\,+\,log\,2, 
\end{equation}
where $F_{0}^{\ast}\,=\,F_0/ \epsilon$ and $\beta\,=\,1/T^\ast$. 

The minimization of the free energy with respect to the variational
parameter gives $S\,=\,\langle \,P_M \,\rangle\,=\,\langle\,P_M\,
\rangle_{Z_1}$ $\;$ i.e. \\
\begin{equation}
 \langle \, P_M \,\rangle
\,=\,\frac{\int_{0}^{1}d(cos\theta)\,P_M(cos\theta)\, 
P(cos\theta)}{\int_{0}^{1}d(cos\theta)\,P(cos\theta)} 
\end{equation}
Solving the above consistency equation one can calculate the energy,
specific heat and orientational order parameters $\langle\,P_M\,\rangle$
for M=2,4 etc. and then the transition can be identified [8].

The cluster variation technique, for improving upon the MF description of
an ordered state was first established by Strieb, Callen and Horwitz
for Heisenberg ferromagnet [9]. Since then, a number of attempts have been 
made to utilize this procedure to explain the disordering transition of
nematics [10-16]. As applied to nematics, this procedure involves an
initial postulate that there exists a mean field which favours orientation
along the director, but terms are added to the free energy expression to
take account in detail of the interactions between particles within what
is called a cluster. For the model potential given by equation (2), the
approximate free energy per particle in the TSC theory is given by [15],
\begin{equation}
 \beta\,F_{2}^{\ast}\,=\,-\,(z/2)\,log\,Z_{12}\,+\,(z-1)\,log\,Z_1 
\end{equation}
where $Z_{12}$ is the two-particle pseudo-partition function \\
\begin{equation}
Z_{12}\,=\,\int \int \,d\omega_{1}\, d\omega_{2} \, exp
\left[\beta\,S\,(z-1)\,\left\{P_4(cos\theta_{1})+P_4(cos\theta_{2})\right\}
\,+\,\beta\, P_4(cos\theta_{12}) \right ] 
\end{equation}
and $d\omega_{i}\,=\,sin\,\theta_{i}\,d\theta_{i}\,d\phi_{i}$. The
condition of minimum free energy with respect to S gives the consistency
requirement 
\begin{equation}
 \langle\,P_4\,\rangle_{Z_1}\,=\,(1/2)\,\langle\,
\left[\,P_4(cos\,\theta_1)+P_4(cos\, \theta_2)\,\right]\,\rangle_{Z_{12}} 
\end{equation}
The various thermodynamic observables are obtainable from the free energy. 
The free energy is given by
\begin{equation}
 U^\ast\,=\,-\,(z/2)\,\sigma_4 ,
\end{equation}
\noindent where $\sigma_L$ are the short-range order parameters 
\begin{equation}
 \sigma_L\,=\,\langle\,P_L(cos\,\theta_{12})\,\rangle_{Z_{12}}
\end{equation}
evaluated at r=a, the nearest neighbour separation.  
The TSC heat capacity can be obtained by differentiating the energy with
respect to the temperature for which exact expressions are given in
referencce [15]. The integrals appearing in the consistency equation were
calculated using the 32-point Gaussian formula and convergence was ensured
by subdivision of the intervals. The consistency point were located by direct
minimization of the free energy in terms of the variational parameter.

The MC simulations reported in this paper have been performed with
periodic boundary conditions on a two dimensional square lattice of
particles interacting through equation (2) for L=4. In particular we have
studied three lattice sizes $10\times10$, $20\times20$ and $30\times30$
using the standard Metropolis 
Monte Carlo algorithm [18-20]. The simulation at the lowest temperature
studied for each of the cases was started from a completely aligned
system, while for the other temperatures, the final configuration of the
equlibration run at the nearby lower temperature was used to start both the
production run at the same temperature and the equlibration run at the
next higher one. 
 Equilibration runs took
between 10000 and 15000 cycles, where one cycle corresponds to N attempted
moves (N being the number of particles in the system). Production
runs were typically between 15000 and 25000 cycles and longer runs, from
50000 to 100000 cycles, have been taken near the heat capacity anomaly for
both the systems. Sub averages of different quantities were calculated
after a certain number of cycles, typically between 1000 and 2000. We have
calculated for each simulation energy, specific heat,  
 second and fourth-rank order parameters and
orientational pair correlation functions.

The second-rank order parameter $\langle\,P_2\,\rangle$ is calculated from
the average over cycles of the largest eigenvalue, $\lambda_{3}$, of the
ordering matrix Q,
\begin{equation}
 Q_{\alpha\beta}\,=\,\langle 
\,q_{i,\alpha} \, q_{i,\beta} \,-\,(1/3)\,\delta_{\alpha,\beta}\,\rangle 
\end{equation}
where the symbols have their usual meaning. The matrix is computed and
diagonalised at every cycle and the average, denoted by the angular
braket, extends to all the particles in the system. The
calculation of the fourth-rank order parameter $\langle\,P_4\,\rangle$
was done according to the algorithm proposed in reference [17].

We have also calculated the second-rank pair correlation
coefficient $\sigma_2(r)$,
\begin{equation}
 \sigma_2(r)\,=\,\langle\,P_2(cos\,\theta_{ij})\,\rangle_r 
\end{equation}
which gives the orientational correlation between two particles i and j
separated by a distance r. This represents the first coefficient in
the expansion of the rotationally invariant angular correlation function
[17-18]. This quantity has been computed for a 80$\times$80 at reduced 
temperatures 0.3, 0.37 and 0.4.

The recently available reweighting technique of Ferrenberg and Swendsen [5] 
has 
been used to accurately predict the order of the phase transition and to locate
the transition temperature in a relatively larger lattice viz a 120$\times$120 
one.
The technique involves generating a vast amount of data to obtain the 
energy-histogram 
at a temperature estimated to be close to the transition temperature and then 
use a transformation to obtain the histogram at neighbouring temperatures. Thus 
the canonical distribution function is directly available at any temperature 
near the transition. To trust the results the parent histogram will have to be 
as accurate as possible and we have used more than one million sweeps of MC 
simulation (per particle) after equilibration has been acheived on the 
120$\times$120 lattice 
at a temperature $T^{\ast}$ = 0.382. The distribution function clearly has a 
two peak structure and by using the reweighting technique we find that the 
two peaks have equal heights at a temperature 0.3827, signalling the precise 
location of the first order transition temperature. The strength of the 
Ferrenberg-Swendsen 
technique lies in the fact that once the distribution function is available 
 at one temperature T, 
 the same at all temperatures in the neighbourhood of T and hence the ensemble
 averages of 
any thermodynamic quantity at these temperatures can be calculated. We have 
calculated $<E>$ in the neighbourhood of transition and from a numerical fit, 
the peak of the derivative, which gives the specific heat $C_v$ (a very sharp 
one indeed), 
has been located at $T^{\ast}$ = 0.3818 strongly suggesting once again a first
 order transition. 

The reduced energy $U^{\ast}$ obtained from MC simulation for the three 
lattice sizes has been plotted as a function of the reduced temperature 
$T^{\ast}$ in figure 1. As is evident from the figure the results nearly 
overlap for all three lattice sizes except predictably near the transition 
region. The prediction of the TSC theory is also depicted in the figure and 
the agreement with the simulation results is reasonably good. We have computed 
the specific heat in two ways for these three lattices; one directly from the 
MC simulation as a fluctuation quantity and the other from a numerical 
differentiation of the energy-temperature curves. The location of the   
specific heat maxima obtained by the two methods agrees well while the peak 
heights do not. A finite size scaling of the transition temperature  obtained 
from the specific heat maximum (including the result for the $120\times120$ 
lattice) 
has been performed and shown in figure 2. The value of $T_c^{\ast} (\infty)$
thus obtained is 0.380.
 In table 1 we present the results along with 
the predictions of the MF and TSC theories. The agreement with the MF result is 
expectedly 
very poor while the TSC results are close to the $T_{c}^{\ast}$ for 10$\times$10 
lattice. This may not be surprising because with decreasing lattice size the 
fluctuations are truncated thus increasing the value of $T_{c}^{\ast}$ and 
bringing it closer to the meanfield predictions [21]. 

Figure 3 shows the results for the order parameters $<P_2>$ and $<P_4>$. The 
changes in both quantities at the transition are sharp, the sharpness 
increasing with 
the lattice size. The TSC and MF predictions are also shown and their 
behaviour 
can again be illustrated by the argument presented in the preceeding paragraph. 
The order parameters show a peculiar behaviour in that, there is a small 
temperature 
region where $<P_4>$ is greater than $<P_2>$. Similar behaviour, it may 
be noted, 
was obtained by [15] for the 3-d $P_4$ model. The MF and the TSC theories 
predicts 
this behaviour and is in qualitative agreement with the MC results.   

Figure 4 shows the variations of $\sigma_2$ and $\sigma_4$, computed for 
the three 
lattice sizes, with temperature along with the MF and TSC predictions. 
The latter 
again is good qualitative agreement with the MC predictions. 

The decay of the 2nd rank correlation function $\sigma_2(r)$ at different 
temperatures 
for r ranging up to half the lattice size (which is 80$\times$80) has been
 shown in 
figure 5. For $T^{\ast}$=0.3 and 0.37 the results could be fitted to a power
 law with 
the decay constant $k_p$ being equal to 0.027 and 0.056 respectively. Above 
the transition, 
namely at $T^{\ast}$=0.4, the decay is clearly exponential with the decay 
constant given by 
0.843. It must be noted that the value of $k_p$ are significantly smaller 
than those 
obtained in the P$_2$-model by Chiccoli, Pasini and Zanoni [3] or that
 predicted by 
the Kosterlitz-Thouless theory [22] for the XY model near the pseudo 
transition temperature.

Finally in figure 6 we show the distribution function obtained for the 
120$\times$120 
lattice at $T^{\ast}$=0.382 and the transformed distribution function at 
$T^{\ast}$=0.3827
 obtained by using the Ferrenberg-Swendsen technique [5]. In the latter 
case the two peaks 
have equal heights signalling the exact transition temperature for this 
lattice size.

We have therefore presented a detailed a Monte Carlo study of the planar 
$P_4$ 
model for various lattice sizes up to L=80 or 120 which seem to be adequete. 
We 
are tempted to conclude that the system undergoes a topological phase 
transition 
which is clearly first order. The exact nature of the topological defects 
leading to 
the phase transition has not been described in the present paper and will 
be reported later. 

\vskip .3in
\centerline {\bf Acknowledgements}
{\small The present calculations were carried out on a DEC ALPHA 3000/600S 
workstation in the department of Physics, Jadavpur University. One of us 
(KM) acknowledges financial support from {\it Council 
of Scientific and Industrial Research}, Government of India.}
\clearpage
\centerline{\bf Table 1}
\begin{center}
\begin{tabular}{|c|c|c|c|c|c|}
\hline
Method & $T_{c}^{\ast}$ & $\langle\,P_2\,\rangle_c$ &
$\langle\,P_4\,\rangle_c$ &  $T_{D}^{\ast}$ & $\Delta S/Nk_B$ \\
\hline
MC  & \, & \, & \,& \, & \, \\
a)\,$10\times 10$  & 0.397 & 0.502 & 0.521 & 0.391 & $\Large -$ \\
b)\,$20\times20$ & 0.386 & 0.410 & 0.404 & 0.381 & $\Large -$ \\
c)\,$30\times30$ & 0.380 & 0.403 & 0.396 & 0.380 & $\Large -$ \\
d)\,$120\times120$ & 0.3818 & $\Large -$ & $\Large -$ & $\Large -$ & 
$\Large -$\\
\hline
MF & 0.501 & 0.430 & 0.561 & $\Large -$ &  1.259 \\
\hline
TSC & 0.404 & 0.503 & 0.604 & $\Large -$ &  1.178 \\
\hline
\end{tabular}
\end{center}
\clearpage
\centerline{\bf Figure Captions}

\noindent 
{\bf Fig.\,1.} \, Energy for the $10\times10$, $20\times20$ and 
$30\times30$
lattices\, :\, 
a)\,(Circles)\, :\,$10\times10$ MC estimates;\, b)\,(Squares)\,
:\, $20\times20$ lattice;\, c) \, (Solid Circles)\, :\, $30\times30$
lattice;\,(d)\,(Continuous line)\, :\, 
TSC results. \\ 
{\bf Fig.\,2.} \, Finite-size scaling of the transition temperatures obtained 
from the specific heat peaks for four lattice sizes from 10$\times$10 to 
120$\times$120. The value of $T_c^{\ast} (\infty)$ is 0.380. \\
{\bf Fig.\,3.} \, Results for the second and fourth-rank long-range order
parameter\,:\,a)\,(Circles)\,:\,$10\times10$ MC
estimates;\,b)\,(Squares)\, :\,$20\times20$ MC results;\, c) \,(Solid
Circles) \,:\,$30\times30$ MC results;\, c)\, (Continuous
line)\,:\,TSC predictions;\,d)\,(Dashed line)\,:\,MF results. \\
{\bf Fig.\,4.} \, Results for the short-range order parameters for a)
$10\times10$, b) $20\times20$ and c) $30\times30$ lattices; 
 \,(Circles)\,:\,MC estimates for
$\sigma_2$;\,(Squares)\, MC estimates for $\sigma_4$; (Continuous
line)\,:\,TSC results 
for $\sigma_2$;\,(Dashed line)\,: TSC results for $\sigma_4$. \\
{\bf Fig.\,5. } \,Second rank correlation function $\sigma_2$ for a) 
$80\times80$ lattice
calculated at
a)\,$T^\ast$\,=\,0.3,\,b)\,$T^\ast$\,=\,0.37;\,c)\,$T^\ast$\,=\,0.4. \\
{\bf Fig.\,6.} \, The probability distribution for energy at $T^{\ast}$ = 
0.3827.
The curve shown has been obtained by reweighting of the distribution function 
originally obtained for $T^{\ast}$ = 0.382. 
\vskip .2in
\centerline{\bf Table Captions}

{\bf Table 1.} \,Transition values of different parameters for $P_4$ model. 
$T_{c}^{\ast}$ 
and $T_{D}^{\ast}$ are respectively the temperatures at which the specific 
heat (obtained from by differentiating the energy-temperature curve) shows 
the maxima and the derivative of the order parameter-temperature curve have 
minima. $<P_2>_c$ and $<P_4>_c$ are respectively the transition values of 
the order parameters. $\Delta S/NK_B$ is the reduced entropy change at the 
transition.
\end{document}